# The appeal of small molecules for practical nonlinear optics

Ivan Biaggio[a]

Dedication: In memory of François Diederich

[a] Prof. Dr. Ivan Biaggio
Department of Physics and Center for Photonics and Nanoelectronics
Lehigh University
Bethlehem, PA 18015, USA
biaggio@lehigh.edu

**Abstract:** Small organic molecules with a π-conjugated system that consists of only a few double or triple bonds can have significantly smaller optical excitation energies when equipped with donor- and acceptor groups, which raises the quantum limits to the molecular polarizabilities. As a consequence, third-order nonlinear optical polarizabilities become orders of magnitude larger than those of molecules of similar size without donor-acceptor substitution. This enables strong third-order nonlinear optical effects (as high as 1000 times those of silica glass) in dense, amorphous monolithic assemblies. These properties, accompanied by the possibility of deposition from the vapor phase and of electric-field poling at higher temperatures, make the resulting materials competitive towards adding an active nonlinear optical or electro-optic functionality to state-of-the-art integrated photonics platforms.

## 1. Introduction

The delocalized electron density in conjugated electronic systems is responsible for the relevant optical properties of organic molecules. By changing the shape of the π-electron system and the substituents attached to it, the physical properties of the molecule, and in particular the way it interacts with light, can be tuned with great flexibility.
These principles can be applied to the design of molecules with a strong nonlinear optical response that can mediate the interaction between photons. Organic materials have therefore always been appreciated for their potential to contribute an active functionality (such as wavelength conversion, electro-optic optic modulation, or all all-optical switching) to many areas of photonics.
This mini-review will discuss the use of small nonlinear optical molecules to build monolithic solid-state organic materials characterized by large nonlinear optical functionalities, with a view towards applications in integrated optics and in modern photonics or nanophotonics platforms. Therefore, one important focus will be the transition from a molecule to a solid-state material and its integration with other integrated optical elements.
While one of the first steps towards the development of a new organic material is the design, synthesis, and characterization of a new molecule, the material science problem of actually using the molecule to build a material with the required properties has not always been addressed, even though it can be a more challenging step than the development of a new molecule. A nonlinear optical organic material, in addition to reflecting the properties of the new molecule, must also have a high optical quality compatible with applications, and its fabrication method must be compatible with existing designs, like the nano-scale waveguides of the silicon-photonics platform.[1]
Newly synthesized molecules, when studied in the solid-state, have often been diluted into a polymer matrix, as a blended, or guest-host system.[2–5] This is a relatively simple way to create a solid state material, but such a solution to the problem of going from molecule to materials intrinsically implies that any advantage that may come from a larger molecular nonlinearity would be frustrated by a low molecular number density. In addition, the fabrication of such polymer-based systems may not always be compatible with modern integrated optics technology.
The ultimate aim of achieving compatibility with integrated optics and photonics circuitry is an important factor in determining the contents of this mini-review, and it also motivates the focus on smaller organic molecules, their monolithic molecular assemblies, and fabrication methods such as physical vapor deposition.
The next section will review the relevant nonlinear optical effects, and highlight the benchmarks that can be used to assess the quality of a molecule, with respect to both how efficiently it achieves a good nonlinear optical response, and its potential to build a high quality nonlinear optical material.
This will be followed by a review of the development of new classes of molecules that have achieved high scores on those benchmarks,[6–14] approached the quantum-limit for the nonlinear optical response,[15,16] could be sublimated without decomposition,[17–19] and enabled molecular beam deposition in vacuum (in addition to spin coating) for the integration with the silicon photonics platform in silicon-organic-hybrid devices.[20–23]

## 2. Nonlinear Optics

The polarization induced in matter by the electric field of an optical wave can have nonlinear components that are proportional to higher powers of the electric field.
A second-order nonlinear optical response leads to the linear electro-optic (Pockels) effect and to the generation of new wavelengths through difference frequency or second harmonic generation. The material property that determines this process is a frequency dependent nonlinear optical susceptibility $\chi^{(2)}_{ijk}(-\omega_3, \omega_2, \omega_1)$.[24,25]
A third-order nonlinear optical response allows the interaction between different optical waves and many other phenomena that can be seen in general as three photons with frequencies $\omega_1$, $\omega_2$, $\omega_3$ interacting to generate a fourth one, with frequency $\omega_4 = \omega_1 + \omega_2 \pm \omega_3$. This process is described by the frequency dependent nonlinear optical susceptibility $\chi^{(3)}_{ijkl}(-\omega_4, \pm\omega_3, \omega_2, \omega_1)$,[24–26] which gives the strength of the third-order response as a function of the frequency and electric field directions of the interacting optical waves.

A second-order nonlinear optical response, like for example the linear electro-optic effect, requires a non-centrosymmetric material, while a third-order response can also be observed in centrosymmetric materials such as amorphous and isotropic molecular assemblies, glasses, or liquids.

The nonlinear optical susceptibilities are in general complex-valued tensors, but this work will mostly use a scalar description for the sake of simplicity.

### 2.1. Organics for nonlinear optics

In organic molecules with extended π-electron conjugation, the wavefunctions of the electronic states that interact with photons occupy relatively large volumes,[27–30] which can lead to large electric dipole matrix elements for optical transitions. In fact, organic dyes are known for their efficiency, and organic materials have led to some of the largest optical nonlinearities observed to date. In addition, the nonlinear optical response of organic molecules is dominated by their electronic states, with little ionic contributions, which makes their off-resonant response practically instantaneous. This applies to the response to an applied electric field for the linear electro-optic effect, and to the speed with which an optical wave can influence another wave when it comes to all-optical switching. This ultrafast capability of organic materials must be contrasted to the ionic response[31] of electro-optic materials like $LiNbO_3$, or to other effects such as carrier injection or excitation in semiconductors, which can lead to an effective nonlinear response, but only at the expense of relatively long relaxation times (due to the lifetime of the injected carriers).

The fast and strong electronic nonlinearities of organic materials, together with the flexibility for tuning their chemical structure towards the optimization of functional properties, make them naturally attractive for various photonics and optoelectronics applications. Examples are third-order effects where the nonlinear optical response mediates the interaction between different optical waves, second-order effects like frequency conversion or parametric down-conversion to create entangled photon pairs, or electro-optic modulators for transferring electronic data streams to the optical domain.

An organic material that presents at the same time a high optical quality, a high nonlinear optical susceptibility, and the ability to use it in integrated optical configurations would be a key enabler for adding the above *active* functionalities to existing *passive*, but extremely advanced, integrated optics platforms, such as silicon photonics.[1,32–34]

### 2.2. From molecules to materials: third-order nonlinear optics

It may be tempting to assume that third-order nonlinear optical interactions are more complex than second-order interactions, and should therefore be addressed later. However, it is actually the other way around: Second-order interactions require the absence of inversion symmetry. This symmetry requirement is an additional factor that determines the nonlinear response of second-order molecules. And even if individual molecules themselves fulfill this requirement, creating a high-quality non-centrosymmetric material from such molecules requires to control their orientation. This is challenging, especially when the material is to be incorporated into existing guided wave technologies.

For third-order interactions, on the other hand, there is no symmetry requirement. This greatly simplifies the discussion of the nonlinearity of single molecules, and of how this molecular property translates into the corresponding nonlinear optical susceptibility of the finished molecular material. The following review of the origin of molecular nonlinearities is based on a fundamental approach that applies to all orders of light-matter interaction, but the main result will be presented with a view towards third-order nonlinear optical interactions.

The two quantities that come into play to describe any nonlinear optical effect are the molecular nonlinearity, represented in this case by the *third-order nonlinear optical polarizability*, γ, and the third-order susceptibility, $\chi^{(3)}$, of the material constructed from those molecules. For the following discussion, both quantities will be taken as *off-resonant*, in the zero-frequency (long wavelength) limit.

The nonlinear optical polarizability relates the applied local electric field and the induced nonlinear dipole on a molecule. The nonlinear optical susceptibility relates the applied electric field to the induced nonlinear optical bulk polarization in a material. In a simple scalar model that neglects interactions between molecules, the third-order susceptibility and the molecular polarizability are related by

$$\chi^{(3)} = f^4 N\,\gamma, \qquad (1)$$

where $f$ is a local field factor that assumes the value $(n^2+2)/3$ in the Lorentz approximation ($n$ is the refractive index in the solid state), and $N$ is the number density of the molecules.

Perturbation theory has been used from the beginning[35] to describe nonlinear optical polarizabilities, which can be expressed as a sum-over-states expansion over all the possible combinations of electric-dipole transitions between states, weighted by the corresponding differences between the transition energies and the energies of the interacting photons.[25,35–38]

This basic approach already allows interesting insights into the structure of the nonlinear optical response, and over the years it has been supplemented by numerous other contributions that discussed various relationships between molecular structure and nonlinear optical response. Examples include bond-length alternation,[39–45] the symmetry of donor-acceptor substitution,[46] the length of π-conjugation,[47] diradical character,[48] the effect of counterions,[49] or modulated conjugation[50] and twisted chromophores.[51,52] For completeness, it is also worth mentioning several studies that focused on the effect of molecular properties such as substitution patterns with donor and acceptor groups,[53–58] the extension of π-conjugation,[59,60] and molecular planarity[58,61] on two photon absorption. (Even though resonant two-photon absorption is not a focus of this review, the same molecular properties affect the off-resonant nonlinearities to be discussed here. In particular, the fundamental quantum limit for resonant two-photon absorption is proportional to the quantum limit for off-resonant quasi-static third-order nonlinearities.[62,63])

The details about the relationships between nonlinear optical response and molecular structure that are listed above are certainly relevant when studying individual molecules, but they may also distract from a more direct assessment of molecular properties based on the most fundamental physical concepts.

One aim of this review is to discuss and use such a global fundamental analysis. This will be done by (1) focusing on experimental data on off-resonant nonlinearities in the zero-frequency limit; and (2) employing physical insights that derive from consideration of transition dipole matrix-elements, excitation energies, the quantum-mechanical limits to the nonlinearity of a molecule, and the benchmarks that can be used to assess the molecular nonlinearity and the potential towards a solid-state material that reflects the molecular properties.

Before discussing the zero-frequency limit, a short side-note about the demonstrated possibilities of designing an essentially off-resonant nonlinear optical response far from it. This can be done by adjusting the location of excitation resonances in molecules where the first optical excitation energy is less than the photon energy.[64,65] In this way it becomes possible to tune the relative value of real and imaginary parts of a nonlinear optical polarizability.[64,65] However, this approach needs to be tailored to specific molecules and photon energies, and it is difficult to discuss it in general terms for different compounds.

The zero-frequency limit for optical nonlinearities can be achieved in practice when the photon energy is significantly smaller (at least by a factor of 2 or more) than the first optical excitation energy. In this limit, all nonlinear optical polarizabilities of the same order become numerically equal,[26] even if they describe different effects, which facilitates the comparison between different compounds. Materials consisting of molecules operating in the zero-frequency limit enjoy a wide transparency range, with low dispersion and a wide-bandwidth ultrafast electronic response. These properties would be of interest for practical applications, in addition to the advantages that stem from a simplified experimental and theoretical approach.

In summary, the focus on the zero-frequency limit allows a better comparison of experimental results and a simplification of the theoretical analysis. Even for applications where higher energy excitation is desired, the optical response in the zero-frequency limit can serve as a general assessment of the efficiency of linear and nonlinear light-matter interaction.

### 2.3. Benchmarks for assessing the third-order nonlinear optical response

The most important contributor to the size of molecular polarizabilities (both linear and nonlinear) is the size of the conjugated system, which is in turn related to the size of the wavefunction of the electrons that interact with a light wave.[27–30] Fundamental quantum physics principles imply that an electron wavefunction that has a larger spatial extent creates a larger dipole matrix element for interacting with photons. In addition, a larger conjugated system normally also leads to a lower energy difference between ground and excited states, which again increases the cross-section for interacting with low-energy photons, as can be immediately seen from the sum-over-states expressions for the polarizabilities.[25,35–38]

Because of its dependence on the size of a molecule, the actual value of a nonlinear optical polarizability is relatively meaningless by itself. Large values associated with big molecules do not necessarily imply a better molecular design. In addition, the corresponding large optical nonlinearities may be attractive at first sight, but are not valuable anymore if those large nonlinearities are defeated by a small number density $N$ in Eq. (1) when molecules must be diluted in a polymer matrix to obtain a solid state material that can be used in applications.

It is therefore much more important to focus on the efficiency of the nonlinear optical polarizability of a given molecule, and the potential of a molecule to produce a solid-state material with the desired qualities.

Basic quantum mechanical principles can be used to determine a fundamental quantum limit to the polarizability of a molecule, which was developed by Kuzyk.[62,66–70] For the off-resonant third-order polarizability, this quantum limit is determined, in S.I. units, by[67,68]

$$\gamma_k = \frac{4}{\epsilon_0}\left[\frac{e^2\hbar^2 N_e}{2m\,E_{10}}\right]^2 \frac{1}{E_{10}^3} = \frac{e^4\hbar^4}{\epsilon_0 m^2}\frac{N_e^2}{E_{10}^5}\,. \qquad (2)$$

Here, the term between the square brackets is an upper limit to the square of the transition matrix element as determined by fundamental sum-rules, while $E_{10}$ is the lowest optical excitation energy. The other quantities in Eq. (2) are the elementary charge $e$, the electron mass $m$, Planck's constant $\hbar$, and the number of electrons in the system, $N_e$, defined as twice the number of multiple bonds in the relevant π-conjugated system when working with organic molecules.[66,67] The actual quantum limit varies between $-\gamma_k$ and $4\gamma_k$, depending if a molecule is centrosymmetric or not,[68] but for consistency in reporting it is better to just use $\gamma_k$ when comparing to actual experimental values, and to not change definitions based on molecular symmetry.

When assessing how efficient a certain molecular design is, one can define an *intrinsic* off-resonant third-order polarizability[71] as the dimensionless ratio

$$\gamma_I = \frac{\gamma_{rot}}{\gamma_k}\,, \qquad (3)$$

where $\gamma_{rot}$ is the experimentally determined orientational average[72] of the off-resonant third-order polarizability in the zero-frequency limit, and $\gamma_k$ is the quantum-limit as defined in Eq. (2).

As mentioned earlier, it is important to evaluate $\gamma_{rot}$ in the zero-frequency limit when using Eq. (3), because in this limit all nonlinear susceptibilities describing different experiments converge towards the same value,[26] and because any misleading enhancements caused by proximity to a resonance are avoided.

Rotational averaging is necessary when nonlinear polarizabilities are measured using molecules in solutions, and it can have a significant effect. As an example, in essentially one-dimensional molecules where a tensor element $\gamma_{1111}$ dominates the third-order polarizability tensor, the value of $\gamma_{rot}$ can be up to 5 times smaller than $\gamma_{1111}$.[72] But if the aim is to create an isotropic material that has polarization-independent nonlinearity because it is made up of molecules with random orientations, then $\gamma_{rot}$ is the right quantity to use.

From Eq. (1) it is clear that one condition to obtain a large third-order susceptibility is to maximize the molecular density, which would happen in a closely packed, single component, monolithic material. As a proxy for the volume occupied by a molecule in such an assembly, it is convenient to use the molecular mass. This leads to the definition of a *specific* third-order polarizability[15],

$$\gamma_S = \frac{\gamma_{rot}}{M}\,, \qquad (4)$$

where $M$ is the mass of the molecule. The specific third-order polarizability defined here allows to estimate the maximum third-order susceptibility that could be obtained in a dense molecular packing with mass density $\rho$ from $\chi^{(3)} = f^4\rho\gamma_S$, which directly follows from Eqs. (1) and (4). A similar alternative figure of merit to assess the effectiveness of the conjugated part of the molecule is to take the ratio between the experimental third-order polarizability and the number of electrons in the conjugated system. This is the nonlinearity per conjugated electron,

$$\gamma_C = \frac{\gamma_{rot}}{N_e}\,. \tag{5}$$

It is important to stress that the figures of merit of Eqs. (3-5) provide valuable *complementary* information. The intrinsic third-order polarizability allows to evaluate the efficiency of the molecule in a scale-invariant way and taking into account the constraints represented by the given energy difference between its electronic states. But it remains possible to modify a molecule in such a way that its excited state energies are lowered, which then *raises the quantum limit itself*, and this can then lead to a significant increase of the nonlinearity per conjugated electron (Eq. 5), or of the *specific* third-order polarizability (Eq. 4). Good molecules should have at least $\gamma_I > 0.01$, $\gamma_C > 0.5\times10^{-48}\,\mathrm{m^5V^{-2}}$, and $\gamma_S > 10^{-23}\,\mathrm{m^5V^{-2}kg^{-1}}$. The latter would deliver a $\chi^{(3)}$ of the order of 1000 times that of fused silica in a material with a refractive index of 2 and a density of the order of that of water.

It is instructive to check the values of the benchmarks of Eqs. (3) and (4) for two simple systems like the helium atom and the hydrogen molecule, whose third-order polarizabilities are $\gamma_{He} = 0.5\times10^{-52}\,\mathrm{m^5V^{-2}}$ and $\gamma_{H2} = 8\times10^{-52}\,\mathrm{m^5V^{-2}}$ (off-resonant values in the static limit, from Ref. [73], after conversion from atomic units, and taking into account a factor of 6 difference in the definition of the third-order polarizabilities). The corresponding dimensionless intrinsic values obtained from $N_e = 2$ and the first excitation energies are $\gamma_I = 0.38$ and $\gamma_I = 0.35$, respectively.[74] These can be compared to one of the best small molecules measured in Ref. [15] and discussed below, TDMEE, which has $N_e = 16$, with an off-resonant $\gamma_{rot} = 8\times10^{-48}\,\mathrm{m^5V^{-2}}$, $\gamma_I = 0.012$, and $\gamma_S = 2\times10^{-23}\,\mathrm{m^5V^{-2}kg^{-1}}$.[74] The nonlinearities of the helium atom and the hydrogen molecule are four orders of magnitude smaller than those of the organic molecule, which is obviously due to the different size of the systems, but the intrinsic values are much more comparable because of their scale invariance.[71] Another observation is that an intrinsic third-order polarizability of the helium atom of ~0.4 is an indirect confirmation of the validity of this quantity as a benchmark for the efficiency of a system. In addition to the intrinsic third-order polarizability mentioned above, the specific third-order susceptibility of Eq. (4) provides a second useful assessment tool for assessing the performance of new molecules when one can assume that it is possible to build a material that consist of a close packing of that molecule, either in a crystal, or in any other kind of dense solid-state assembly.

### 2.4. From molecules to materials: second-order nonlinear optics and electro-optics

The major challenge of assembling any optimized nonlinear optical molecule into a solid state material that reflects the molecular properties has been mentioned above. This general difficulty is exacerbated for second-order nonlinear optical effects, which require symmetry breaking and an orientational molecular order in the material: The development of organic molecular materials for second-order nonlinear optics and electro-optics has to address all the issues that were discussed above for third-order nonlinear optics, plus the challenge of having to control molecular orientation to achieve the required noncentrosymmetric order. In addition, this review aims at maintaining a singular focus on organic molecular materials that are compatible with state-of-the-art photonics platforms and integrated optics circuitry, where the challenge of obtaining a non-centrosymmetric molecular order is magnified even more.

Crystal growth has for years been the natural method to obtain second-order nonlinear optical activity and electro-optic effects in macroscopic molecular materials,[75] but similar techniques have found more limited use in integrated optics applications.[76–80] Some molecular glasses are also characterized by some intrinsic noncentrosymmetric order when vapor deposited,[81–85] but were not developed for second-order nonlinear optics or electro-optics. Other examples of relevant systems are mixtures of nonlinear optical molecules and liquid-crystal-like molecules, which can show a second-order nonlinear optical activity when poled by an electric field,[86] or self-assembled films with an orientational order governed by directional hydrogen bonding.[87,88]

In general, the most widespread, most utilized method to create organic electro-optic materials in a flexible way has relied on the principle of first creating a material that contains randomly oriented molecules, and then inducing an average orientational order by applying an electric field at higher temperatures. This is the route taken in many systems where strong nonlinear optical molecules are diluted in a polymer matrix.[89,90]

In fact, electrically poled polymers have been in development for 30 years and have made great progress. While they have not found commercial applications yet, they are currently the organic materials that have shown the highest electro-optic effects, and they have also already been used to demonstrate electro-optic modulation on the silicon photonics platform.[90–93]

## 3. Organics for practical nonlinear optics

From the point of view of practical applications, organic materials do have some limitations. They are typically not well adapted to working with high power lasers and will likely never be a competition for established second-order inorganic nonlinear optical crystals used for wavelength conversion, Pockels cells, or electro-optic modulation. The same applies to third-order nonlinear optical applications at higher intensities, even though organics have the advantage of off-resonant third-order nonlinear optical susceptibilities that can be orders of magnitude larger than inorganic materials with similar refractive indices.

However, organics can have real benefits for applications that need large nonlinear optical efficiencies, when working with smaller intensities, or when interaction lengths are limited. This is particular relevant for applications in integrated optics, where continuing progress in the design and fabrication of optical circuitry at the nanoscale is enabling unprecedented control of optical modes and their propagation on a chip. This is seen, in particular, in the nanoscale, silicon-based optical circuitry of the silicon photonics platform.[1,32,94]

The combination of the flexibility of molecular design and chemical synthesis with existing integrated optics technology merges the best of two worlds: the very accurate guided wave technology enabled by silicon-on-insulator technology, and the flexibility of molecular design and synthesis. This is the basic idea behind silicon-organic-hybrid devices, as it is exemplified by the slot-waveguide, two silicon rails that guide a an electro-magnetic mode propagating between them[1,32,34,94–96] (see also Figure 5, below).

To contribute to any of these applications, organic molecular materials must score highly in the benchmarks discussed above (Eqs. 3-5), and must also be easy to integrate with modern nano-scale photonic devices.

This mini-review next focuses on the results of a collaboration between Prof. Diederich's group at ETHZ and Lehigh University. The aim of the research was to provide new answers to a deceptively simple question: how can one create a solid-state organic material that has a high optical quality (low losses, low scattering), has a large nonlinear optical response, and can be efficiently integrated into state-of-the-art nanoscale photonic structures?

When looking for a solution, we first concentrated on third-order nonlinear optics to study the molecular nonlinearities separately from the problem of symmetry, and consciously looked for a path towards materials that would be complementary to existing approaches. In practice, this meant ignoring the well-established polymers and guest-host systems, and trying instead to obtain the required nonlinearities and optical quality using dense assemblies of a single molecule. Such monolithic systems can be a viable and flexible alternative route towards the application of organic nonlinear optics in integrated optics. They are potentially well described by Eq. (1), implying the necessity of optimizing both the molecular third-order polarizability and the molecular density, as represented by the benchmarks of Eqs. (3) and (4).

The next section is dedicated to a review of the small molecules with donor-acceptor substitution that fulfilled these requirements.

### 3.1. Donor-acceptor substitution

This section reviews research that cycled between the synthesis of new molecules,[6–13,22,46,97–100] their nonlinear optical characterization,[10,15,16,19,63,101] and ultimately the development of a solid-state material for photonics.[1,17,18,20,96,102]

The first step was the extension of a basic tetraethynylethene scaffold via nitrogen substitution and the addition of other groups to obtain cyanoethynylethene-based molecules.[7] This then led do a set of new compact donor-acceptor substituted molecules.[8,9] One of them (TDMEE) is shown in Figure 1.

The main message from the characterization of the optical and nonlinear optical properties of these molecules[15,16] was this: the lowest optical excitation energy (related to the energy gap between highest occupied molecular orbital, HOMO, and lowest unoccupied molecular orbital, LUMO) only decreased weakly as the conjugated system was extended, and most of the molecular variations had similar intrinsic third-order polarizabilities (Eq. 3), close to the highest values ever reported.[15] It turns out that both observations are related to each other: These molecules have donor and acceptor groups substituted around the basic cyanoethynylethene structure (DA substitution). Strong electron acceptors (CN groups) are coupled to strong N,N-dimethylanilino electron donors, often via an ethynylene bridge (triple bond), which has been shown to be a suitable inert conjugated spacer between the dimethylanilino donor and the tricyanovinyl acceptor. The effect of DA substitution, when compared to unsubstituted molecules of similar size, is a HOMO-LUMO gap that is *lower* in absolute terms and decreases much more slowly as the size of the conjugated system increases (see later below for more details about this).

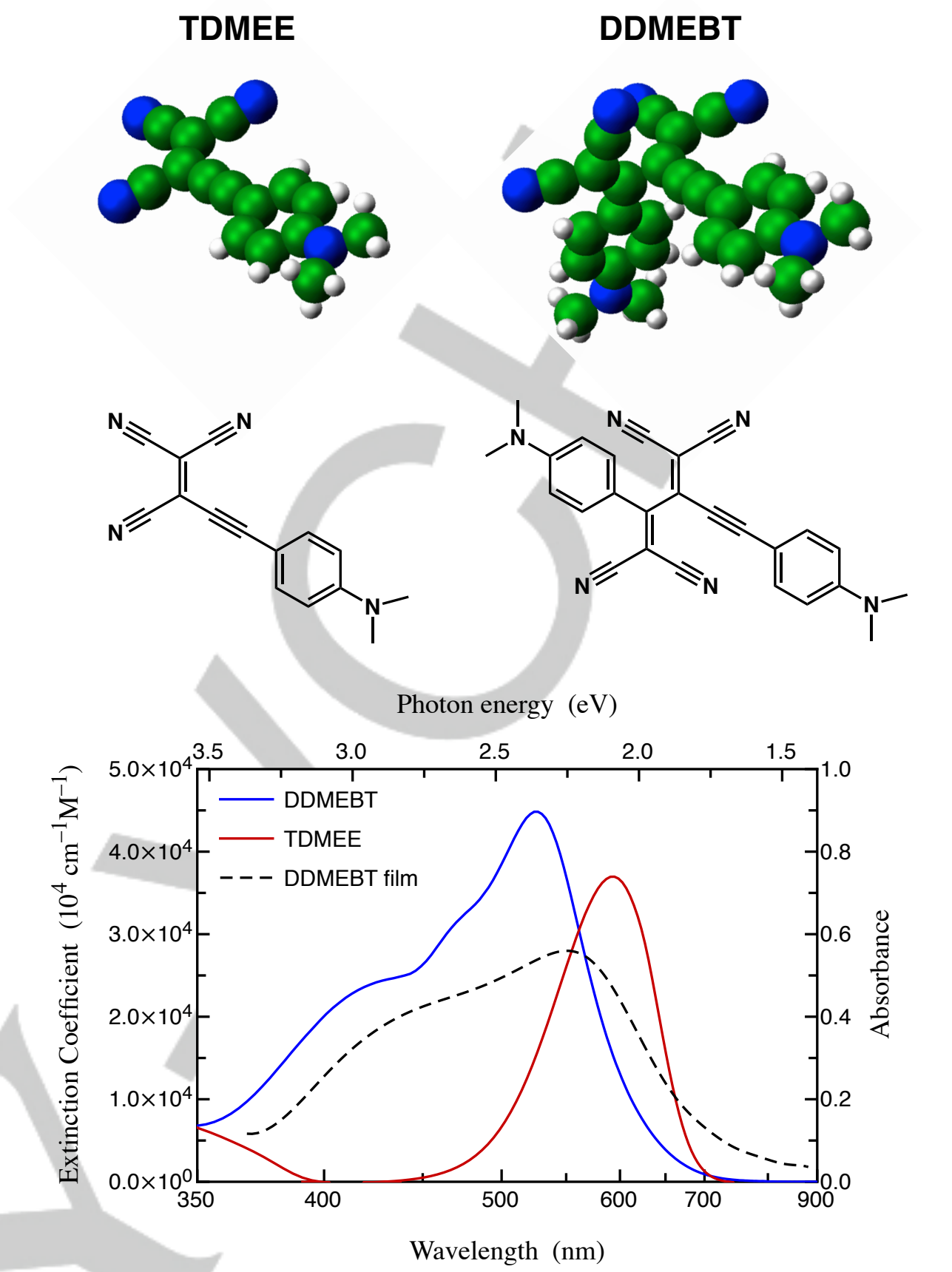

**Figure 1.** 3D models and chemical structures of two representative DA-substituted cyanoethynylethenes (TDMEE, Refs. [8,9]) and tetracyanobutadienes (DDMEBT, Refs. [10,11]). The graph plots the extinction coefficient when the two molecules are dissolved in dichloromethane (solid curves, referring to left axis), and the absorbance of a ~70 nm solid-state film of DDMEBT (dashed curve, right axis). Note that the curves are plotted as a function of photon energy (top horizontal axis), with the lower horizontal axis giving the corresponding wavelength. The DDMEBT solid state film was chosen to be very thin in order to avoid deformations of the spectrum that would arise from multiple reflections inside the film and their interference.

For molecules with similar transition dipole matrix element between ground and excited state, a smaller HOMO-LUMO gap directly leads to a larger off-resonant third-order polarizability, as easily seen in sum-over-state expansions.[25,35,36,38] The strong dependence of the third-order polarizability from the excited state energies is also evident from the fact that the quantum-limit to the third-order polarizability is inversely proportional to the *fifth power* of the first optical excitation energy. Since lower excitation energies raise the fundamental quantum limit for the third-order polarizability, the DA-substituted cyanoethynylethene molecules,[15,16] which have similar intrinsic third-order polarizabilities, achieve nonlinearities orders of magnitude larger than other molecules of similar size without DA substitution.[103–105] This same basic property was maintained when extending the same principle to DA-substituted tetracyanobutadiene molecules.[10,11]

For the sake of brevity, this mini-review is highlighting two molecules, that are representative of the two molecular families introduced above. Their chemical structure and related data are shown in Figure 1, and their properties are summarized in Table 1. When comparing these two molecules, one sees that DDMEBT contains almost the complete TDMEE molecule as a sub-unit. Therefore, it is not surprising that the nonlinearity for DDMEBT ($\gamma_{rot}$ in Table 1) is similar to TDMEE. But in Table 1 one also sees that the specific third-order polarizability is clearly smaller for DDMEBT, reflecting the fact that it is not planar, with two separate conjugated systems, only one of them large enough to contribute significantly to its nonlinearity. Finally DDMEBT has an advantage with its intrinsic third-order polarizability value, but only because this value was calculated using only the number of electrons in the main conjugated system. It should also be noted that the definition for the intrinsic third-order polarizability $\gamma_I$ (see Eq. 2) depends on the rotationally averaged experimental value, which is 5 times lower than the one for an optimal light-polarization in molecules that have a dominant direction of the nonlinearity.[72] In terms of the proximity of these molecules to the fundamental limit, one can say that their third-order polarizability is roughly 10 times lower. One other extremely important property of the DDMEBT molecule that is not revealed just by the figure or merit in Table 1 is the fact that the its non-planarity allows it to condense into an essentially amorphous solid state, as will be discussed in much more detail in the next section.

**Table 1.** Properties of the two molecules of Figure 1

| | TDMEE | DDMEBT[b] |
|---|---|---|
| Molar mass (g/mol) | 246.3 | 416.47 |
| $\lambda_{max}$ (nm) | 591 | 527 |
| $E_{10}$ (eV) | 2.10 | 2.35 |
| $N_e$ (dimensionless) | 16 | 14[a] |
| $\gamma_k$ ( $10^{-48}$m$^5$V$^{-2}$) | 661 | 286 |
| $\gamma_{rot}$ ( $10^{-48}$m$^5$V$^{-2}$) | 8 | 6 |
| $\gamma_C$ ( $10^{-48}$m$^5$V$^{-2}$) | 0.53 | 0.43 |
| $\gamma_S$ (m$^5$V$^{-2}$kg$^{-1}$) | 1.95 | 0.87[a] |
| $\gamma_I$ (dimensionless) | 0.012 | 0.021 |

[a] This molecule has a broken conjugation; the value here is for the main conjugated system. [b] This value is calculated with the mass of the full molecule, to better reflect the practical nature of this figure of merit.

In addition to reaching a highly efficient third-order polarizability, the above work on the DA-substituted cyanoethynylethenes also demonstrated several other interesting properties of DA molecules. In a nutshell, the conjugated system that separates donor and acceptor groups allows some level of communication between them, and more efficient conjugated pathways of similar size generally lead to an *increase* in the HOMO-LUMO gap, which is the opposite of what generally happens in non DA-substituted systems.[8,106,107] When the donor and acceptor groups become separated by a large conjugated system, the communication between them is reduced. As a consequence, the HOMO-LUMO gap still decreases when the conjugated system grows in size, but not as steeply as for non-substituted molecules.

Another important issue is that, as the donor and acceptor groups separate, the ground-sate probability density (HOMO) will tend to be more concentrated near the donor, while it is the other way around for the excited state (LUMO). Because of this, the overlap integral between ground and excited states, which determines the dipole transition matrix element, is expected to start decreasing as the donor and acceptor groups separate.[12,107,108] This decrease will then work against a large polarizability.

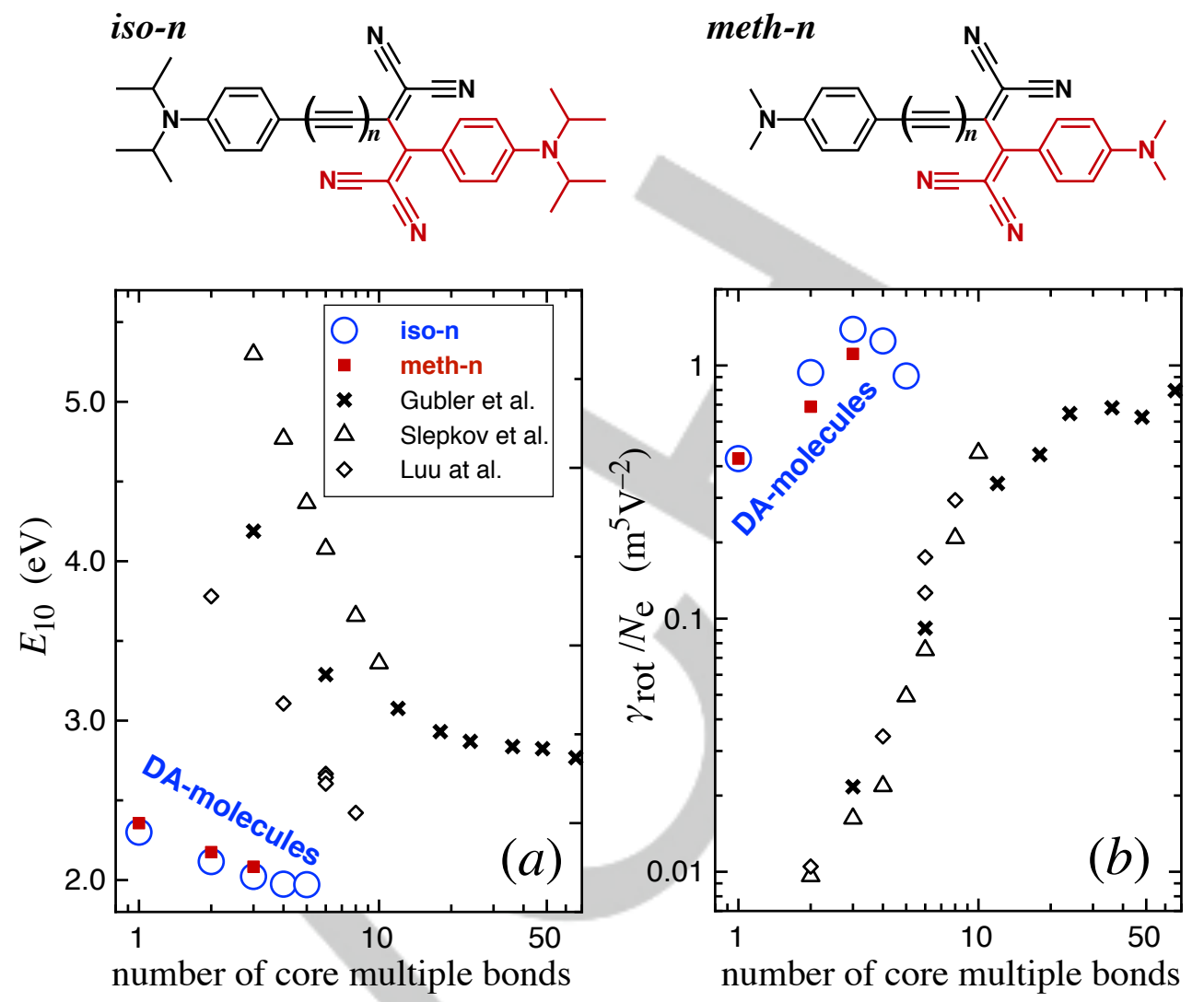

**Figure 2.** First optical excitation energy (*a*) and rotational average of the third-order polarizability per conjugated electron (*b*) in the DA-substituted molecules shown at the top, measured in dichloromethane solutions. Both quantities are represented as a function of the size of the conjugated system separating them, expressed as the number of core multiple bonds. The DA-substituted molecules, shown as coloured data points in the plots, are compared to several other molecules which lack DA-substitution, and where the extension of the conjugated system was systematically varied,[103–105], shown as different data point symbols, in black. Adapted from Ref. [63].

The above argument implies that DA substitution, while effective in creating large molecular polarizabilities in the small molecules described above, is expected to stop working for larger molecules. A first confirmation of this effect for a systematic extension of a linear conjugated system was obtained in Ref. [109] despite some issues with instability and weak solubility of the compounds. A second stronger confirmation came thanks to new compounds partially based on the DDMEBT end-groups,[14] which could then be fully characterized from the point of view of nonlinear optics in Ref. [63].

One of the main results of this study is shown in Figure 2. The DA-substituted molecules that follow the basic design of DDMEBT maintain a first optical excitation energy (and therefore a HOMO-LUMO gap) that is generally by a factor of ~2 smaller than similar molecules without substitution (from Refs. [103–105]). Partly because of this, the DA-substituted molecules have a third-order polarizability that is about *two orders of magnitude larger* than the comparison molecules. Another result of this study was that DA substitution can still lead to improvements in the molecular third-order nonlinearity for a linear conjugated system consisting of up to 3 multiple bonds, before the negative effect of DA substitution on the transition dipole matrix element sets in and the third-order polarizability starts dropping (see Figure 2). The non-substituted molecules, on the other hand, have a third-order nonlinearity that shows a continued steep increase up to a conjugated systems with 10 multiple bonds, but they never actually surpass the efficiency of the DA-substituted molecules, as represented by the nonlinearity per multiple bond $\gamma_C$ (Eq. 5).

The trend for the first optical excitation energy, $E_{10}$, displayed in Figure 2(a), is a general feature of small DA-substituted molecules that has also been observed in Ref. [16]: while the HOMO-LUMO gap did not change between molecules that only differed by the size of the conjugated system, the third-order polarizability did increase by more than a factor of two.[16] This superlinear increase is consistent with the fact that the molecules remain close to the fundamental quantum limit of Eq. (2), which is proportional to $N_e^2$ (recall that $N_e$ is the number of electrons in the conjugated system, a proxy for its size). The steeper power laws seen in other molecular systems without DA-substitution,[103–105] whose nonlinearities are shown in Figure 2, are caused by the concurrent increase in transition dipole moment and decrease in HOMO-LUMO gap.

This data about the slower increase of optical nonlinearity with size in DA-substituted molecules should not be interpreted as a lack of effectiveness in these molecules. Rather, it should be interpreted as the ability, enabled by the presence of donors and acceptor, to keep the nonlinearity high while the size of the molecule shrinks.

All of the above demonstrates that push-pull systems, long a mainstay of molecules for second-order nonlinear optics, can be extremely important for third-order nonlinear optical effects in small molecules.

It is important to stress that the claim here is not that the effect of donors and acceptors on the third-order nonlinearities of organic molecules has not been studied earlier. As already mentioned in the introduction, the positive effect of donors and acceptors has been analyzed by several investigators in the past. The claim here is that DA-substitution, applied in a targeted way to small molecules, can have a dramatic positive effect on the molecular nonlinearities that can be achieved. In small molecules, DA-substitution can be used to set the HOMO-LUMO gap and increase the third-order susceptibility in a way that can be well explained and analyzed using the sum-over-states expansion of the third-order polarizability[25,35–38] and the fundamental quantum limit developed by Kuzyk (Eq. 2).

### 3.2. Vapor deposition for third-order nonlinear optics

The effectiveness of DA substitution in raising the third-order nonlinearity of small molecules to a level that makes them competitive with molecules three or more times larger implies the possibility of greatly enhanced third-order susceptibilities (estimated to be at least three orders of magnitude larger than for silica glass) if this kind of molecules can be combined into a dense solid-state. Such a material could be a crystal, or an amorphous, dense supramolecular assembly that resembles a molecular glass. The latter choice seems to be the only good alternative. Crystal growth would be difficult to implement, and would be counterproductive if it resulted in an assembly of micro-crystalline grains that would lead to unacceptable scattering losses. Dilution into polymers must be excluded because it would also dilute the optical nonlinearity.

Since the field of application for such molecular materials would ideally be integrated optics, towards such applications as silicon-organic-hybrid devices embedded in optical circuitry,[1,33,34,79,94] the use of a flexible and efficient fabrication method is required. Deposition of amorphous materials from solution, for example via spin casting, would in principle allow the formation of the desired dense supramolecular assemblies,[23] but the reliance on a solvent would be a drawback that could make it hard to effectively deposit a homogenous molecular assembly inside nano-scale devices like silicon photonic crystal structures, or a silicon slot waveguide, where silicon rails have a trench with a width of 100 nm or less between them.

The molecules discussed above have a small mass and a record-high nonlinear optical response, but in addition to that they can also sublimate without decomposition. Molecular beam deposition in vacuum, or vapor deposition in general, then becomes a very good possibility for creating the finished molecular assembly. This deposition in vacuum would be a robust and flexible method to homogenously fill any nano-sized gaps on previously created nanoscale circuitry.

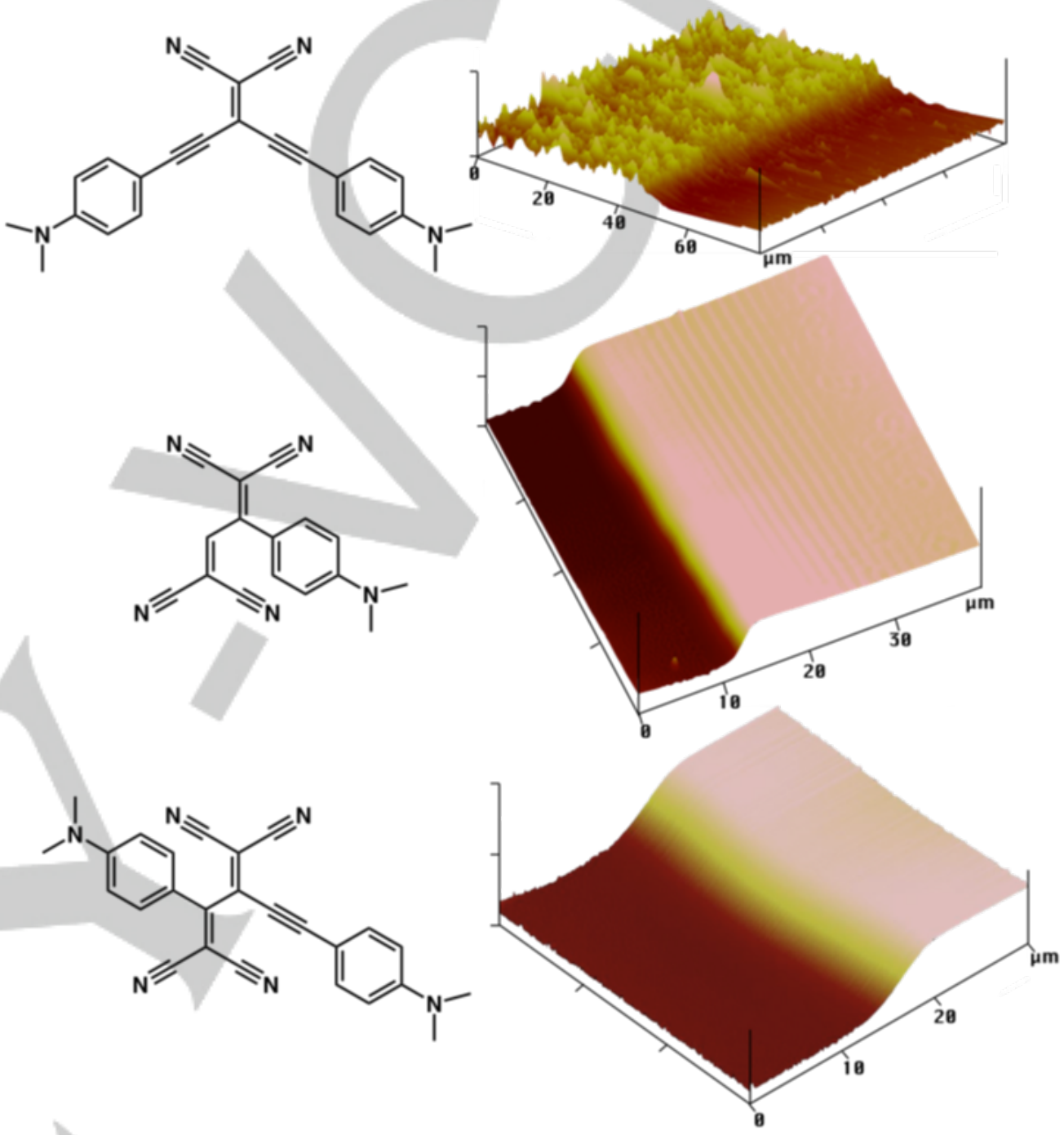

**Figure 3.** Atomic force microscope (AFM) scans of films obtained by molecular beam deposition of three different molecules on a glass substrate. From top to bottom: 0.5 µm thick film from a cyanoethynylethene molecule,[106] 1.6 µm thick film of a small tetracyanobutadiene molecule,[11] and 1.0 µm thick film of the larger tetracyanobutadiene DDMEBT (see Figure 1). For each film, the AFM scans were performed near a section of the film that was protected from the molecular beam via a mask, and the AFM plots show the transition from the substrate to the organic's film surface.

In summary, vapor deposition of small molecules, combined with the large specific nonlinearities of DA-substituted molecules, could be an ideal tool to create the organic component of hybrid integrated optics devices. Molecular beam deposition allows the use of masks to confine the deposition to any selected small area, has a nanometer-level control of thickness, is free from solvents, and can even be used to create multilayer systems (solvent-based deposition, on the other hand, would affect any existing organic film). An example of the potential of molecular beam and vapor deposition of small organic molecules are the nowadays ubiquitous display screens based on organic light emitting diodes (OLEDs), which are vapor deposited through masks to form the pixel grid.

A difficulty from the point of view of material development is that the necessity of being compatible with vapor-deposition and of forming an essentially amorphous material upon deposition adds a new requirement to the design of a new molecule. In fact, not all molecules that sublimate without decomposition can form a high-quality material upon deposition from the vapor phase. Strong intermolecular interaction can lead to crystal growth, which would then result in a material consisting of microcrystals, which can lead to large scattering losses.

As an example, Figure 3 shows atomic force microscope pictures of films formed by organic molecular beam deposition of a cyanoethynylethene compound from Ref. [106], and two tetracyanobutadiene compounds from Ref. [11], one of them is the DDMEBT molecule introduced in Figure 1. The cyanoethynylethene compound is planar, like TDMEE in Figure 1, and it forms multicrystalline films with unacceptable light scattering. The tetracyanobutadiene compounds, on the other hand, form very homogenous films with a surface flatness of the order of a few nanometers. Therefore, while TDMEE has a larger specific third-order polarizability, promising potentially a larger third-order susceptibility in the solid state, the intermolecular interactions are too strong, leading to the formation of nano- or micro-crystals. From the point of view of pure efficiency, DDMEBT has a low specific third-order polarizability and has a broken conjugation. But it also has a non-planar structure that leads to an essentially amorphous film upon deposition.

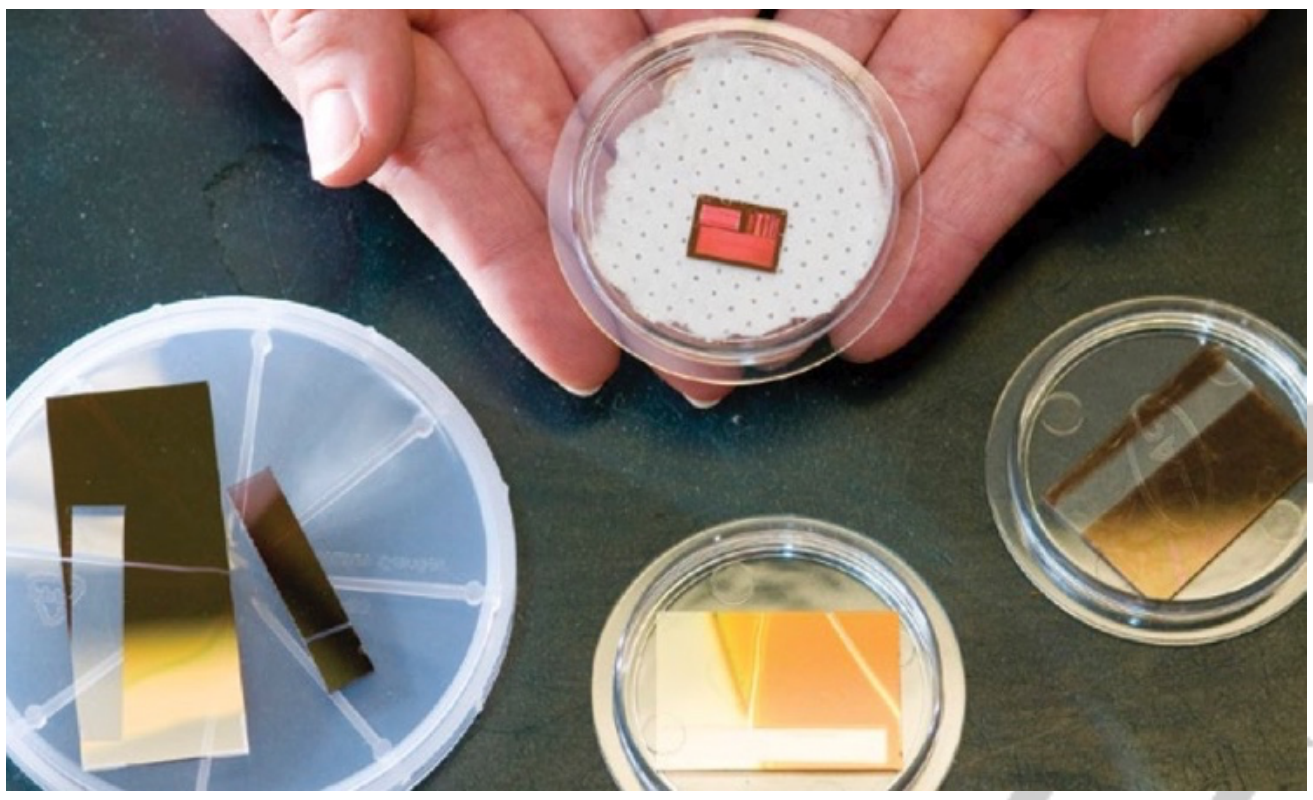

**Figure 4.** Photograph of various DDMEBT samples of varying thickness, on glass slides and on a silicon microchip.

More specifically, there is no space for all the substituent groups in the tetracyanobutadienes of Ref. [11] to be parallel to the same plane: they have to choose between an s-cis and an s-trans conformation by rotating around the single bond. While the second molecule in Figure 3 has an s-trans conformation, DDMEBT has a rotation around its single bond by an angle of 96.7 degrees,[11] close to perpendicular. It is therefore possible that it can take both conformations when subjected to thermal perturbations. This flexibility during the film deposition process could aid towards forming an amorphous solid state, and may also play a role when poling the DDMEBT solid state by application of an electric field at higher temperatures (see below).

The high quality of the vapor-deposited DDMEBT material has been confirmed by the absence of light scattering and by optical spectroscopy of thin film transmission, which showed well-defined oscillations caused by multiple-reflection interference.[17] The same experiments also determined the material's refractive index dispersion.[17] The off-resonant third-order susceptibility of the material was measured to be $\chi^{(3)}_{1111} = (1.0 \pm 0.3) \times 10^{-19}\ \mathrm{m^2V^{-2}}$ in off-resonant third harmonic generation experiments[19] and $\chi^{(3)}_{1111} = (2 \pm 1) \times 10^{-19}\ \mathrm{m^2V^{-2}}$ in off-resonant degenerate four-wave mixing experiments,[17] with a corresponding nonlinear refractive index $n_2 = (1.7 \pm 0.8) \times 10^{-19}\mathrm{cm^2W^{-1}}$ (which determines a refractive index change $n_2 I$ in the presence of an intensity $I$). These off resonant values were measured at a fundamental wavelength of 1.5 µm, where the refractive index of the material is 1.8,[17] and they are three order of magnitude larger than for silica glass with a refractive index of 1.5. The low refractive index of 1.8 is significantly less than the refractive index of silicon, which is important for integration with silicon-on-oxide waveguides because it allows the optical field to extend into the organic material.

Since the material is amorphous and since there is no requirement for lattice matching, vapor depositing DDMEBT on any substrate leads to a homogenous supramolecular assembly with very high optical quality. Figure 4 is a photograph of a few as-grown DDMEBT films. Visually, thick DDMEBT films look similar to gold films. AFM characterization shows that they have a smooth surface with a surface roughness below a few nm.[22]

All these attractive material properties make DDMEBT a good candidate for building a silicon-organic-hybrid device.[20] To achieve this, the material was deposited on a slot waveguide via organic moleclar beam deposition in high vacuum. After deposition, the waveguide was sliced both perpendicular and parallel to the light propagation direction using a focused ion beam, and scanning electron microscope pictures of the organic filling were taken. This was done at several different positions, and no grains or voids were found.[20] Figure 5 shows a cartoon representing the basic idea of using molecular beam deposition to fill the nanoscale structure of the slot waveguide, and the scanning electron microscope pictures that were obtained.

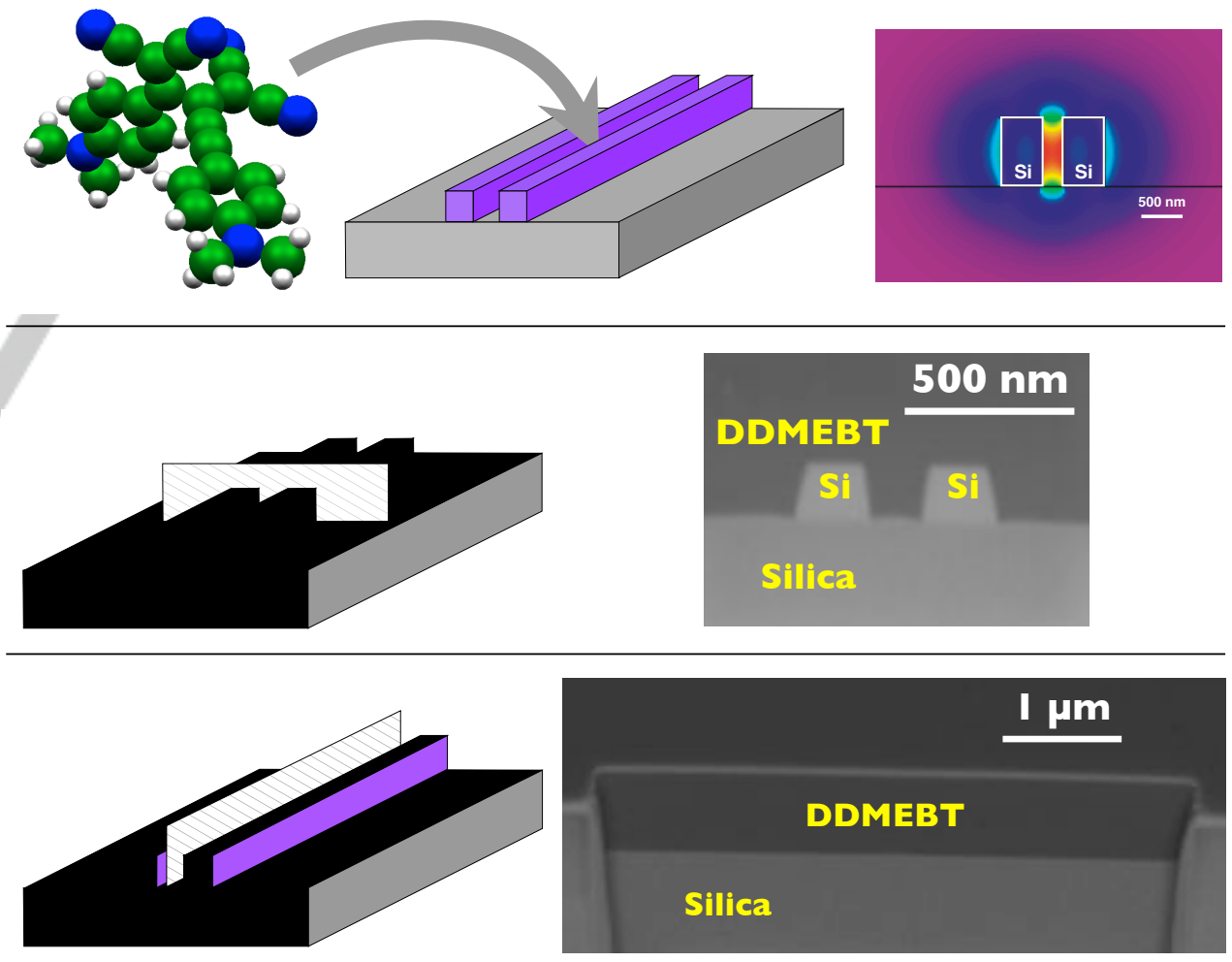

**Figure 5.** Top: Cartoon of the basic idea of vapor deposition or molecular beam deposition in high vacuum, with DDMEBT molecules individually entering the space between two silicon rails that define a slot waveguide, therefore covering it up and homogeneously filling the slot in the middle. The optical mode propagates inside this slot thanks to the low refractive index of the organic filling, as shown by the rightmost figure that plots the spatial distribution of the intensity. Middle and bottom: scanning electron microscope pictures of the homogeneity of the molecular material inside the slot, both along a cross-section of the waveguide and along the direction of propagation of the light.

The silicon-organic-hybrid slot waveguide based on DDMEBT was used in a demonstration of all-optical switching where a 170.8 Gbit/s data stream was optically demultiplexed down to 42.7 Gb/s.[20] It was observed in this experiment that the organic material did not have an influence on the losses of the millimeter-long waveguide that was at the core of the all optical switching scheme.[20] Further characterization of DDMEBT-based silicon-organic hybrid waveguides also demonstrated the expected femtosecond switching-time enabled by the off-resonant electronic nonlinearity of the organic filling.[96] Finally, while vapor deposition is still the preferred choice, it was later demonstrated that good DDMEBT films can also be obtained via spin coating and subsequent baking to remove nanoscopic and microscopic solvent bubbles.[23]

### 3.3. Vapor deposition for second-order nonlinear optics

Section 2.4 highlighted a main difficulty that must be faced when creating a second-order nonlinear optical material, the requirement of a non-centrosymmetric molecular order. In polymer-based materials, such an order is most often achieved by higher-temperature electric field poling.[90–93] The question then arises if the same technique could be applied to the small-molecule, vapor-deposited monolithic assemblies described above.

Since the DDMEBT molecule has already been demonstrated as a high-quality organic component in silicon-organic-hybrid devices, and since, as a DA-substituted molecule, it also has both a second-order nonlinear optical polarizability and a molecular dipole moment (equal to 11 Debye[99]), it makes sense to address the question of using electric-field poling to impart an electro-optic functionality to such monolithic small-molecule assemblies. Electrically poled, vapor-deposited organic materials would represent a new and flexible tool for adding electro-optic functionality to photonics platforms.

The DDMEBT assembly appears to have properties similar to a class of a vapor-deposited organic glasses, where a slight preferential molecular orientation is imparted by the vapor deposition process itself.[83–85,110,111] Such a material may be expected to have a glass-transition temperature above which molecular reorientation and partial alignment could become possible under application of an electric field.

In fact, it has been recently shown that the DDMEBT molecular assembly can indeed be electrically poled above 80°C,[102] and that the procedure leads to a relatively high electro-optic coefficient of 20 pm/V[102] that remains stable at room temperature. This value is already equal to two thirds of the largest electro-optic coefficient of $LiNbO_3$, and could in principle already be used in a silicon-organic-hybrid electro-optic modulator based on a slot waveguide.[112]

This result is both surprising and interesting. One reason is that molecules in a dense supramolecular assembly should not as easily rotate in an applied electric field as when molecules are embedded in a polymer matrix (where higher temperature poling is normally used). In a dense assembly each molecule would have to rely on the displacement and rotation of its neighbors in order for a macroscopic preferential order to be established. However, there are indications that DDMEBT, with its non-planar structure that consists of two planes almost normal to each other,[11] and on the verge between an s-cis and an s-trans configuration,[11] has potential to adapt to its environment, especially at higher temperatures. It is therefore possible that the DDMEBT material behaves like an organic molecular glass above the glass transition temperature, and that it reconfigures as new orientations of the constituent molecules are established, leading to the observed high electro-optic coefficient of 20 pm/V.[102] The promise of these results is that they have been obtained with a molecule that was originally designed for third-order effects, and that has been in no way optimized for this application, both from the point of view of its second-order polarizability, and from the point of view of controlling its ability to reorient in the solid state, to facilitate electric-field poling. There is likely much room for improvement towards the development of monolithic vapor-deposited small molecule assemblies for second-order nonlinear optics.

## 4. Conclusions

This mini-review described the principles behind new molecular families that systematically approached the quantum limit for third-order nonlinear optics to within a factor of ~10 (a number that may be a fundamental characteristic of carbon-based chemistry), while at the same time having an exceptionally large specific third-order polarizability. The fundamental reason for this can be found in the use of donor and acceptor groups around a compact π-conjugated electron system.

Molecules designed according to this principle were used to build monolithic solid-state assemblies that are amorphous and have a high optical quality, in addition to possessing a large optical nonlinearity. The small molecular mass of these compounds allows sublimation without decomposition, so that organic nonlinear optical materials can be efficiently added to modern integrated photonics devices and nanostructures by clean and controllable vapor deposition.

The molecule that was the main result of all these developments is DDMEBT, first reported in Refs. [10,17]. Its vapor-deposited material was used as the organic nonlinear optical component in a silicon-organic-hybrid device for ultrafast all-optical switching,[96] and it has also been recently used to demonstrate higher-temperature poling of its molecular assembly to impart an electro-optic response with an electro-optic coefficient of 20 pm/V.[102]

The DDMEBT material has now become one of the reference materials for both experimental and theoretical work in the literature. It can be expected that additional investigations using this compound, or other compounds derived from the same basic idea, will lead to further progress in the future.

Prof. Biaggio obtained his PhD from the Swiss Federal Institute of Technology (ETH) in Zürich, was a postdoc with Prof. Hellwarth at USC, with whom he kept up a collaboration on large polarons, and has been a physics professor at Lehigh University since 2002, where he is working on nonlinear optics, solid-state physics, and laser spectroscopy, with a standing interest in organic materials for photonics and organic semiconductors for optoelectronics applications.
He is an Optica Fellow.

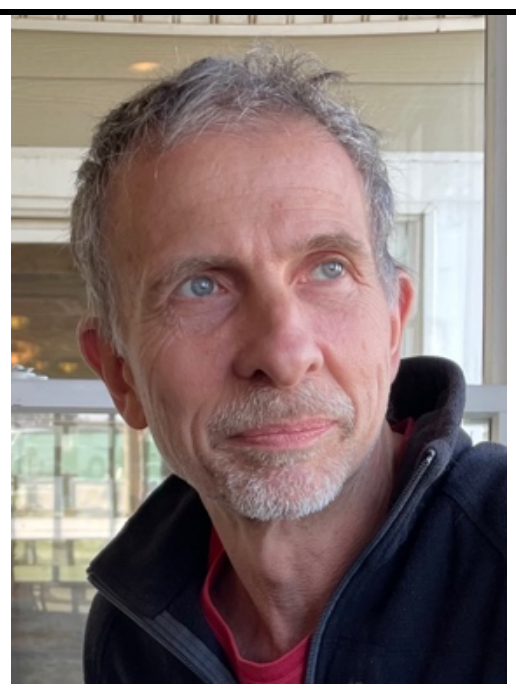

## Entry for the Table of Contents

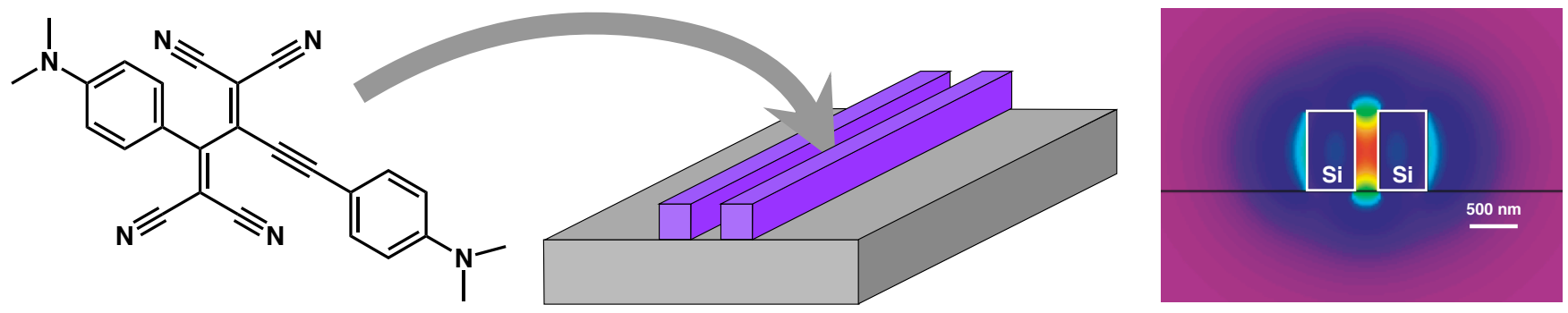

Donor-acceptor substitution in small molecules with a compact π-conjugated system leads to record-high molecular nonlinearities in small sizes. A dense monolithic assembly of such molecules can have extraordinarily strong nonlinear optical properties. The ability to vapor deposit such small molecule materials makes them highly competitive for adding all-optical switching and/or electro-optic functionality to state-of-the-art integrated photonics.